\documentclass[sigconf]{acmart}
\usepackage{multirow}
\usepackage{float}
\usepackage{stfloats}
\usepackage{array}
\usepackage{makecell}
\usepackage{enumitem}

\usepackage{adjustbox}
\AtBeginDocument{%
  }

\copyrightyear{2025}
\acmYear{2025}
\setcopyright{acmlicensed}\acmConference[WWW Companion '25]{Companion Proceedings of the ACM Web Conference 2025}{April 28-May 2, 2025}{Sydney, NSW, Australia}
\acmBooktitle{Companion Proceedings of the ACM Web Conference 2025 (WWW Companion '25), April 28-May 2, 2025, Sydney, NSW, Australia}
\acmDOI{10.1145/3701716.3715258}
\acmISBN{979-8-4007-1331-6/2025/04}

\begin{document}

\title[RecUserSim: A Realistic and Diverse User Simulator for Evaluating Conversational \\ Recommender Systems]{RecUserSim: A Realistic and Diverse User Simulator for Evaluating Conversational Recommender Systems}

\author{Luyu Chen}
\authornote{Co-first authors}
\authornote{Work under the internship in Huawei Noah’s Ark Lab}
\affiliation{
\institution{Gaoling School of Artificial Intelligence, Renmin University of China
\city{Beijing}
\country{China}}}
\email{luyu.chen@ruc.edu.cn}

\author{Quanyu Dai}
\authornotemark[1]
\affiliation{%
  \institution{Huawei Noah's Ark Lab}
  \city{Shenzhen}
  \country{China}
}
\email{daiquanyu@huawei.com}

\author{Zeyu Zhang}
\affiliation{
\institution{Gaoling School of Artificial Intelligence, Renmin University of China
\city{Beijing}
\country{China}}}
\email{zeyuzhang@ruc.edu.cn}

\author{Xueyang Feng}
\affiliation{
\institution{Gaoling School of Artificial Intelligence, Renmin University of China
\city{Beijing}
\country{China}}}
\email{xueyangfeng@ruc.edu.cn}

\author{Mingyu Zhang}
\affiliation{%
  \institution{Huawei Technologies Ltd.}
  \city{Shenzhen}
  \country{China}
}
\email{zhangmingyu14@huawei.com}

\author{Pengcheng Tang}
\affiliation{%
  \institution{Huawei Technologies Ltd.}
  \city{Shenzhen}
  \country{China}
}
\email{tangpengcheng12@huawei.com}

\author{Xu Chen}
\authornote{Corresponding author}
\affiliation{
\institution{Gaoling School of Artificial Intelligence, Renmin University of China
\city{Beijing}
\country{China}}}
\email{xu.chen@ruc.edu.cn}

\author{Yue Zhu}
\affiliation{%
  \institution{Huawei Technologies Ltd.}
  \city{Shenzhen}
  \country{China}
}
\email{zhuyue9@huawei.com}

\author{Zhenhua Dong}
\affiliation{%
  \institution{Huawei Noah's Ark Lab}
  \city{Shenzhen}
  \country{China}
}
\email{dongzhenhua@huawei.com}

\renewcommand{\shortauthors}{Luyu Chen et al.}

\begin{abstract}
Conversational recommender systems (CRS) enhance user experience through multi-turn interactions, yet evaluating CRS remains challenging. User simulators can provide comprehensive evaluations through interactions with CRS, but building realistic and diverse simulators is difficult. While recent work leverages large language models (LLMs) to simulate user interactions, they still fall short in emulating individual real users across diverse scenarios and lack explicit rating mechanisms for quantitative evaluation. To address these gaps, we propose RecUserSim, an LLM agent-based user simulator with enhanced simulation realism and diversity while providing explicit scores. 
RecUserSim features several key modules: a profile module for defining realistic and diverse user personas, a memory module for tracking interaction history and discovering unknown preferences, and a core action module inspired by Bounded Rationality theory that enables nuanced decision-making while generating more fine-grained actions and personalized responses. To further enhance output control, a refinement module is designed to fine-tune final responses.
Experiments demonstrate that RecUserSim generates diverse, controllable outputs and produces realistic, high-quality dialogues, even with smaller base LLMs. The ratings generated by RecUserSim show high consistency across different base LLMs, highlighting its effectiveness for CRS evaluation.
\end{abstract}





\maketitle

\section{Introduction} \label{cha:intro}

\textcolor{black}{Conversational recommender system (CRS) can interact with users through natural language to deliver personalized items, products, or services based on their preferences and feedback~\cite{10.1145/3453154,GAO2021100}. 
It has numerous potential applications that can significantly impact people's lives, such as e-commerce~\cite{liu-etal-2023-ecommerce}, streaming services~\cite{gao2023chat}, and travel booking~\cite{liao2019deep}.
Compared to traditional recommender systems, which often operate passively in the background based on static user profile or past behavior, it offers a user-centered experience by actively engaging users in multi-turn interactive conversations and gathering real-time feedback.
However, building and evaluating a CRS remains challenging due to its dynamic and interactive nature~\cite{GAO2021100,wang-etal-2023-rethinking-evaluation}. In this work, we focus on the evaluation of CRS.}

Traditional evaluation methods focus on assessing recommendation accuracy and response quality in isolation by comparing them with the ``ground-truth'' conversations from fixed benchmark~\cite{BARCOR,KGSF}. 
Such a protocol significantly underestimates the importance of the dynamic interaction process, failing to effectively capture the true performance of a CRS~\cite{wang-etal-2023-rethinking-evaluation}.
Online user testing is considered the gold standard for CRS evaluation, as it captures real user interactions and feedback. However, its high time consumption and substantial labor costs make it difficult to scale~\cite{GAO2021100}.

To tackle this problem, user simulators have emerged as a practical alternative~\cite{2020/Evaluating_CRS_via_User_Simulation,10.1145/UserSimCRS,2018/Conversational_Recommender_System,2020/Graph_for_Conversational_Recommendation,2020/Towards_Question-based_Recommender_Systems}. 
\textcolor{black}{Ideally, an effective user simulator should accurately model individual user behavior given a specific persona, enabling realistic role-playing, while also being able to capture the diversity of large-scale user population. These capabilities are essential for precise and comprehensive CRS evaluation~\cite{2020/Evaluating_CRS_via_User_Simulation,10.1145/UserSimCRS}. }
However, achieving both objectives presents significant challenges. \textcolor{black}{Traditional rule-based simulators fail to realistically mimic user behavior in a dynamic conversations due to their lack of flexibility and adaptability~\cite{GAO2021100,2020/Evaluating_CRS_via_User_Simulation,10.1145/UserSimCRS}}.

Recent advancements in large language models have given them strong capabilities in dialogue understanding, text generation, and role-playing~\cite{LLMSurvey,2023/UncoveringGPT,survey-RPLAs}, making them a compelling alternative to traditional approaches. 
Building on these advancements, LLM-based user simulators have emerged, including single-prompt methods~\cite{wang-etal-2023-rethinking-evaluation,zhu2024llmbasedcontrollablescalablehumaninvolved,kim-etal-2024-pearl} and agent-based architectures~\cite{2024/reliableUser}.
However, current approaches remain far from sufficient in developing an ideal user simulator.
First, they fail to perform fine-grained simulation of individual user behavior, including language, actions, and decision-making processes. Second, they struggle to capture the diversity of real user population in terms of language styles and action types. They often generate uniform language styles (determined by the base LLM) and execute only a limited set of actions (constrained by either rigid prompting templates~\cite{wang-etal-2023-rethinking-evaluation,kim-etal-2024-pearl} or restricted action space~\cite{2024/reliableUser}). These limitations primarily stem from their reliance on fixed benchmark datasets with coarse-grained user profile and weak role-playing capabilities, making it difficult to model nuanced user behavior and diverse interaction patterns.

\textcolor{black}{
To overcome the above limitations, we propose \textbf{RecUserSim}, an LLM-based autonomous agent for user simulation in conversational recommender systems.
}
RecUserSim achieves both realistic user role-playing and diverse user population representation through a carefully designed agent framework comprising four key modules: profile module, memory module, action module, and refinement module.
Specifically, \textcolor{black}{
the profile module, serving as the foundation of the simulator, constructs fine-grained and diverse user profile. It is further enhanced with a conflict resolution mechanism to ensure realistic and coherent user representation.}
\textcolor{black}{
The memory module tracks user preferences and interaction history to ensure consistent behavior simulation. An unknown preference excitation mechanism is designed to dynamically uncover users' latent interests.
The action module, as the core component of RecUserSim, \textcolor{black}{is inspired by the Bounded Rationality theory~\cite{simon1955behavioral}, which conceptualizes decision-making as a three-step process: receiving and processing information, evaluating options, and making decisions. We adapt this theory into our three-tier ‘Rating-Action-Response’ mechanism to better mimic real user behavior.}
The refinement module enhances RecUserSim’s control over output language by adjusting its information richness, formality, and sentence length to align with the user’s persona.
Together, these modules ensure the simulator adheres to the fine-grained user persona while enabling a comprehensive representation of the diverse user population.
}

In summary, this paper makes the following contributions:

\begin{itemize}[leftmargin=10pt, itemsep=2pt, parsep=0pt, topsep=2pt]
    \item We propose a novel LLM agent-based user simulator, \textbf{RecUserSim}, which enables both realistic individual role-playing and diverse user population representation for the accurate and comprehensive evaluation of conversational recommender systems.
    
    \item \textcolor{black}{We design several key mechanisms to enhance RecUserSim's role-playing capabilities, such as the `Rating-Action-Response' mechanism for more realistic behavior modeling and the tool-augmented refinement method to ensure persona adherence.}
    
    \item \textcolor{black}{We validate RecUserSim’s superiority in both individual user simulation and diverse user population representation through comparative analyses with existing models. We further demonstrate the robustness and reliability of RecUserSim’s rating mechanism in capturing CRS performance across different base LLMs. Finally, we deploy RecUserSim to develop and evaluate Huawei’s Celia Food Assistant, highlighting its industrial applicability.}

\end{itemize}

\section{Related Work}
Conversational recommender systems utilize multi-turn dialogue to dynamically adapt recommendations to user needs. 
Unlike the evaluation of traditional recommender systems, which can focus solely on recommendation accuracy based on explicit ratings and user-item interactions~\cite{hasan2024based,2001/Item-basedRS,schafer2007collaborative,2016/Wide&Deep}, CRS requires simultaneous assessment of recommendation accuracy, dialogue quality, and system interactivity.
Consequently, developing comprehensive and effective CRS evaluation methods remains a key challenge~\cite{GAO2021100}.

\subsection{Metric-based Evaluation Methods}
Traditional CRS evaluation approaches often focus on turn-level metrics, where recommendation accuracy and dialogue quality are assessed separately. For recommendation evaluation, metrics such as hit rate, precision, recall, F1 score, and NDCG~\cite{10.1145/NDCG} are commonly used. Dalogue quality is typically assessed using automated metrics like BLEU~\cite{10.3115/1073083.1073135} and ROUGE~\cite{lin-2004-rouge}. However, these metrics rely heavily on predefined reference in dialogue datasets, limiting their effectiveness since user feedback in real conversations often deviates from these references~\cite{GAO2021100,jannach2020end,2023/RSbias}.

\vspace{-1pt}
\subsection{Simulator-based Evaluation Methods}
To better align with the interactive nature of CRS~\cite{2020/Estimation-Action-Reflection,lei2020interactive}, simulator-based evaluation methods have been proposed to simulate the complete dialogue process between the user and the system~\cite{GAO2021100}.

\textbf{Traditional User Simulators.} \textcolor{black}{Human evaluation can assess CRS performance in multi-turn dialogues, but its high cost limits large-scale application~\cite{2018/OpenSearch,2019/Off-Policy_Evaluation,2015/From_Infrastructure_to_Cultur}}. Alternatively, researchers proposed rule-based and template-designed user simulators~\cite{GAO2021100,2020/Evaluating_CRS_via_User_Simulation,10.1145/UserSimCRS,2018/Conversational_Recommender_System,2020/Graph_for_Conversational_Recommendation,2020/Towards_Question-based_Recommender_Systems} to enable CRS to conduct complete conversations. However, these simulators struggle to capture the flexibility and adaptability of real users, limiting their effectiveness~\cite{GAO2021100,2020/Evaluating_CRS_via_User_Simulation,10.1145/UserSimCRS}.

\textbf{LLM-based User Simulators.} \textcolor{black}{Recently, researchers have begun leveraging LLMs’ powerful capabilities to develop more realistic user simulators for CRS evaluation~\cite{wang-etal-2023-rethinking-evaluation,zhu2024llmbasedcontrollablescalablehumaninvolved,kim-etal-2024-pearl}. Single-prompt LLM simulators, such as iEvalLM~\cite{wang-etal-2023-rethinking-evaluation}, MACRS~\cite{fang2024multi}, and PEARL~\cite{kim-etal-2024-pearl}, have been proposed to assess CRS performance. However, these simulators suffer from limited dialogue diversity, repetitive conversation flows, and insufficient control over generated outputs due to their reliance on single prompt templates~\cite{2024/reliableUser}.
}
\textcolor{black}{To address these limitations, ~\citet{2024/reliableUser} proposed CSHI, which introduced LLM-based agents to user simulation with basic user profile and action space. While CSHI marks progress, it struggles to capture diverse user behaviors and lacks precise control over generated outputs, making it difficult to produce realistic, persona-consistent interactions, thereby limiting its effectiveness in user simulation.}

In contrast, our method overcomes these limitations by incorporating a richer set of user personas and an expanded action space with flexible and diverse options, significantly enhancing user simulation diversity. 
Furthermore, the tool-augmented refinement module provides fine-grained control over output language, while the `Rating-Action-Response' mechanism replicates real users’ decision-making processes, enabling realistic individual role-playing.

\begin{figure*}[t]
  \centering
  \vspace{-1.5pt}
  \includegraphics[width=\linewidth]{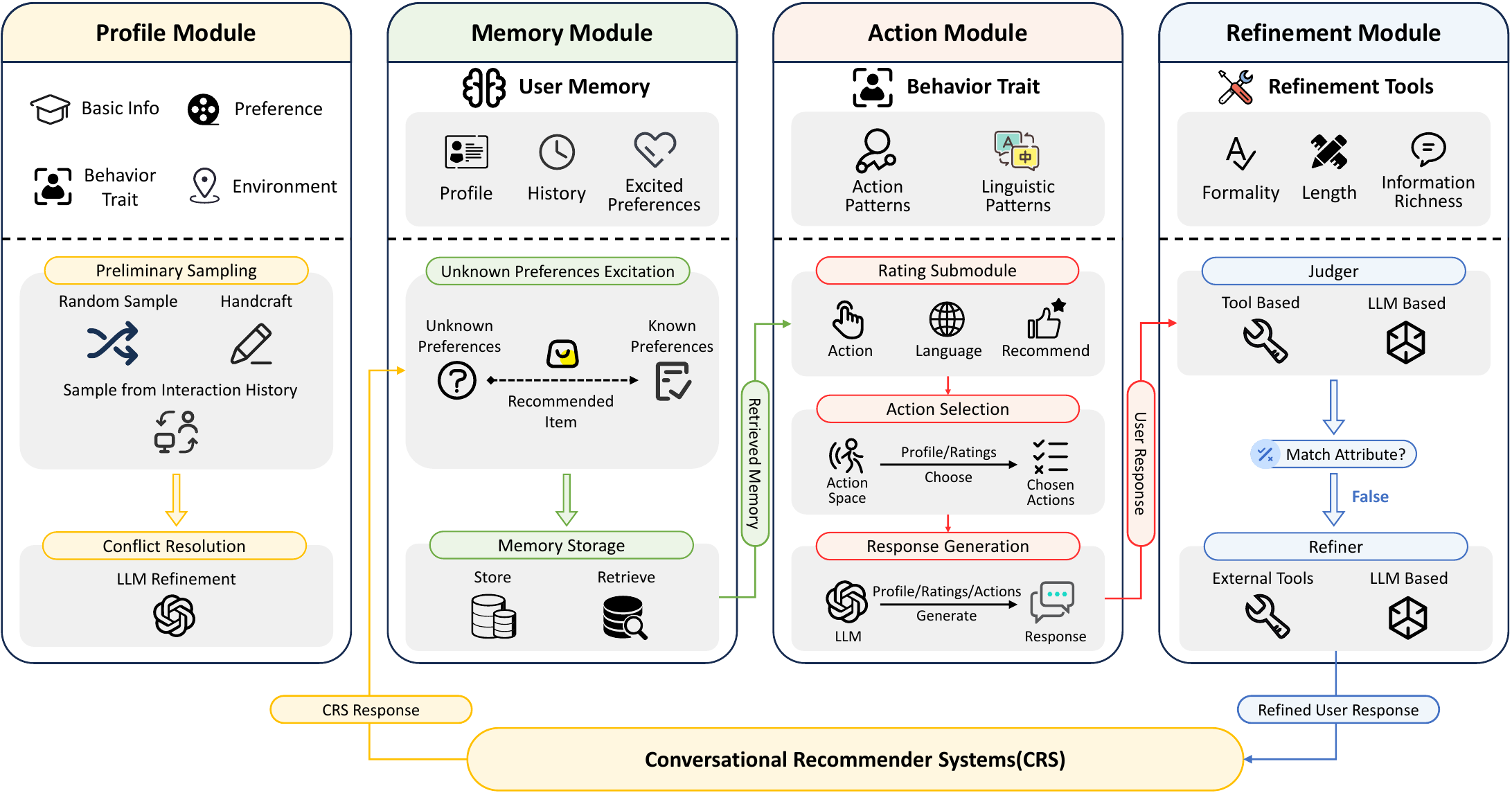}
  \caption{An overview of RecUserSim, which consists of four modules: profile, memory, action, and refinement. The profile module constructs user's profile and resolves its conflicts. This profile is stored in the memory module along with the interaction history. Based on this information, the action module scores the CRS's response, selects appropriate actions, and generates responses. Finally, the refinement module fine-tunes these responses to align with user's linguistic patterns.}
  \label{model structure}
  \vspace{-8pt}
\end{figure*}

\vspace{1.5pt}
\section{Methodology}
\vspace{1.5pt}
\subsection{Overview}
\vspace{0.5pt}
In this paper, we propose a novel user simulator based on LLM-based autonomous agent, namely \textbf{RecUserSim}, for evaluating conversational recommender systems.
Figure~\ref{model structure} provides an overview of RecUserSim, which comprises four main modules. Together, these modules enable our user simulator to perform fine-grained user role-playing and represent a diverse user population.

The \textbf{profile module} constructs detailed user personas, providing a strong foundation for simulation realism and diversity. By employing a conflict resolution mechanism, it addresses inherent inconsistencies in user profile, ensuring the generation of realistic and varied user representations. \textcolor{black}{The \textbf{memory module} stores user profile while tracking interaction history, ensuring behavioral consistency. It incorporates an unknown preference excitation mechanism to dynamically uncover latent user interests, enabling the simulator to better mimic real user preferences.} The \textbf{action module}, consisting of three submodules, mimics a real user's decision-making process. Through multi-dimensional rating, \textcolor{black}{fine-grained} action selection, and \textcolor{black}{personalized} response generation, this three-tier action mechanism results in more realistic and diverse user responses. Finally, given that the responses generated by the action module may not fully align with the user's \textcolor{black}{linguistic patterns}, we employ a tool-augmented \textbf{refinement module}. This ensures that output \textcolor{black}{language} closely matches the predefined user \textcolor{black}{persona}, providing nuanced control over the simulator's responses.

\vspace{-3pt}
\subsection{Profile and Memory Management}

\subsubsection{Profile module}
An effective user simulator should accurately represent a wide range of individuals to reflect real-world user variability~\cite{MemSim}. To achieve this, our simulator incorporates a \textcolor{black}{fine-grained} user profile, further enhanced by a conflict resolution mechanism, to ensure the generation of \textcolor{black}{realistic} and diverse user profile.

\textbf{Diverse Profile Construction.} The user profile consists of four key aspects: \textit{basic information}, \textit{environment information}, \textit{preferences}, and \textit{behavior traits}, allowing for nuanced user modeling. \textit{Behavior traits} encompass user's \textcolor{black}{linguistic and action patterns}, which shape user's communication styles and influence user's decision-making process, leading to diverse actions and responses. \textcolor{black}{To ensure diversity, we construct user profile by randomly sampling from predefined dictionaries according to prior distributions, rather than manually designing them or deriving them from interaction history.}

\textbf{Profile Conflict Resolution.}
Random sampling may inadvertently produce conflicting attribute combinations—for instance, a user who dislikes spicy food but prefers Sichuan cuisine (which is typically spicy). \textcolor{black}{To ensure coherence and realism, we leverage LLM to assess sampled profile and adjust conflicting attributes.} This process enhances simulation quality by preventing illogical or contradictory user characteristics.

\subsubsection{Memory module}

The memory module is a pivotal component that bridges the profile and action modules for LLM-based agents~\cite{AgentSurvey,MemSurvey}. It stores the user profile sampled by the profile module and tracks past interactions. This supports the subsequent action module in rating, action selection, and response generation.

\textcolor{black}{\textbf{Unknown Preference Excitation.}
Real users often discover new interests over time, making it essential to simulate unknown preferences for realistic user modeling. Existing approaches, such as~\cite{zhu2024llmbasedcontrollablescalablehumaninvolved}, attempt to achieve this by hiding a subset of known preferences from the user model and revealing them when a matching recommendation appears. However, this method assumes all unknown preferences are predefined rather than genuinely latent, overlooking the fact that users may not always be aware of their own preferences~\cite{2014/unknown_preferences}.}

\textcolor{black}{To address this, we introduce an LLM-driven unknown preference excitation mechanism that dynamically uncovers latent interests. Specifically, if a recommended item is not among the user's known preferences but is highly agreeable (e.g., aligns with their taste), the LLM recognizes it as a new preference. This adaptive updating process enables the memory module to track evolving user interests, enhancing interaction realism.}

\subsection{Three-Tier Action Mechanism} \label{cha:action_mecha}
\textcolor{black}{An ideal user simulator should accurately model individual user behavior. To achieve this, we draw inspiration from the Bounded Rationality Model in economics~\cite{simon1955behavioral} and propose a \textbf{Tree-Tier Action Mechanism}, which better captures user decision-making in CRS interactions.}

\textcolor{black}{In economics, the Bounded Rationality Model abstracts user behavior into: receiving and processing information, evaluating options, and making decisions. 
Applied to recommender systems, when receiving a recommendation, the user first process the given information (receiving and processing Information). Next, they evaluate the recommendation (evaluating options). Finally, they select an appropriate action, deciding whether to accept or reject the recommendation (making decisions), before conveying their decision and any further requests to the CRS through natural language.}

To formalize this process, we introduce a \textbf{Tree-Tier Action Mechanism}: `\textit{Rating-Action-Response}', which enables more realistic and diverse user interactions. Upon receiving a CRS response, RecUserSim applies the \textbf{Multi-Dimensional Rating} mechanism, generating quantitative evaluations that provide signals for refining CRS recommendations. Based on these evaluations, the \textbf{Fine-Grained Action Selection} mechanism determines the simulator’s next actions, ensuring realistic and diverse behaviors. Finally, the \textbf{Personalized Response Generation} mechanism integrates user-specific linguistic patterns and attitudes toward the CRS, producing responses aligned with individual speaking styles and preferences.

\textbf{Multi-Dimensional Rating.}~\textcolor{black}{This module simulates users' evaluation processes, providing a more accurate assessment of CRS performance while actively guiding the subsequent action selection and response generation.} \textcolor{black}{Inspired by Generative Verifier~\cite{zhang2024generative}, we enhance rating reliability by first prompting the LLM to generate justifications before assigning scores.} Ratings are structured across three key dimensions as suggested by~\cite{CRSEval}:
\begin{itemize}[leftmargin=10pt, itemsep=2pt, parsep=0pt, topsep=2pt]
    \item \textit{Language quality}: \textcolor{black}{Reflects how well the generated dialogue aligns with human language in terms of naturalness, fluency, and clarity. Rated from 1 to 5.}

    \item \textit{Action quality}: \textcolor{black}{Assesses whether the CRS selects the correct action and understands the user's request. For example, if a user asks for a spicy restaurant, the CRS should take the `recommendation' action (correct action selection) and provide a list of spicy restaurants (accurate intent comprehension). The rating (1–5) is determined by how well these two criteria are met.}

    \item \textit{Recommendation quality}: \textcolor{black}{Applies only when the CRS performs `recommendation' action. The final score is the sum of objective score (1–5) which measures how well the recommendation matches user preferences, and subjective modifiers (each ranging from -1 to +1) that adjusts for user-specific tendencies. For instance, a restaurant that moderately aligns with user preferences may receive a base score of 3. If the user enjoys exploring new cuisines and the restaurant is novel, the score may increase by 1, resulting in a final rating of 4.}
\end{itemize}

\textbf{Fine-Grained Action Selection.}~Traditional single-prompt-based simulators often exhibit uncontrollable behavior patterns. CSHI~\cite{zhu2024llmbasedcontrollablescalablehumaninvolved} attempts to address this by introducing a basic action selection module that simulates user decision-making with three predefined actions. However, this approach results in rigid and uniform user behaviors, lacking personalization. To overcome these limitations, we significantly expand the action space and enable more flexible, fine-grained action selection.

Our framework defines five user actions and allows multiple actions to be selected simultaneously, better reflecting real user behavior. For example, when receiving an unsatisfactory recommendation, a user might provide negative feedback while clarifying preferences at the same time. This multi-action capability enhances interaction diversity and flexibility. Additionally, our design is highly extensible, allowing new actions to be added or removed based on specific needs. For instance, a "chit-chat" action could be introduced to support broader research objectives.

The five user action types in our framework are:

\begin{itemize}[leftmargin=10pt, itemsep=2pt, parsep=0pt, topsep=2pt]
\item \textit{Request for recommendations}: The user actively seeks recommendations, usually at the start of a conversation.
\item \textit{Preference clarification}: The user clarifies their preferences to help the CRS \textcolor{black}{generate better-aligned recommendations, espsecially after receiving an unsuitable suggestion.}
\item \textit{Feedback on recommendation}: The user provides positive or negative feedback on a recommendation, guiding the CRS in refining future suggestions.
\item \textit{Item attribute inquiry}: The user requests additional details about a recommended item, like location or price.
\item \textit{End conversation}: The user terminates the interaction, either after receiving a satisfactory recommendation or due to repeated unsatisfactory results.
\end{itemize}

Furthermore, our simulator allows users with different action patterns to exhibit distinct tendencies in action selection, thereby enhancing flexibility and personalization. Even under similar preference and recommendation scenarios, users with different action patterns may choose different actions. For instance, a casual user might end the conversation after receiving a satisfactory recommendation, whereas a more indecisive user may further inquire about additional details.

\textbf{Personalized Response Generation.}~
The response generation submodule produces natural language feedback from users to the CRS. Previous methods, which rely solely on dialogue history and selected actions, often overlook user-specific language styles and attitudes toward the system, making the simulator unable to accurately mimic an individual user's response.

To address this, we leverage LLMs to generate personalized responses by integrating user profile, dialogue history, satisfaction ratings, and selected actions. Since LLMs struggle to directly associate numerical values with satisfaction levels, we first convert satisfaction scores into descriptive text, enabling the model to better interpret user attitudes. Additionally, we embed users' linguistic patterns into LLM prompts, ensuring responses align with individual speaking styles.

However, when multiple constraints are applied simultaneously, LLMs may fail to fully satisfy all requirements. To mitigate this, we introduce a refinement module that fine-tunes responses, ensuring they better align with all specified linguistic patterns.

\subsection{Tool-augmented Refinement}
\textcolor{black}{Large language models often struggle to balance multiple output constraints~\cite{2024/LLM_output_CONTROL}. For instance, an LLM tasked with generating a response that is concise, informal, and information-rich may  struggle to satisfy all criteria simultaneously.} To address this, our user simulator employs a refinement module that applies constraint-specific adjustments sequentially, ensuring outputs align more closely with predefined user personas.

The refinement module consists of several specialized refinement tools, each fine-tuning outputs based on specific linguistic patterns including information richness, formality, and sentence length. \textcolor{black}{Each tool includes a judger to assess alignment and a refiner to modify the output if necessary.} Rule-based judgers can handle straightforward patterns like sentence length, while more complex aspects like formality rely on LLMs for both judger and refiner.

In scenarios where a user ends a conversation, strict adherence to linguistic patterns may not be necessary. As a result, language refinement tools can be deactivated for such specific actions.

We design refinement tools for three linguistic patterns:

\begin{itemize}[leftmargin=10pt, itemsep=2pt, parsep=0pt, topsep=2pt] 

    \item \textcolor{black}{\textit{Information richness}: The tool uses an LLM to identify key points in a sentence (e.g., time, location, preferences) and count their occurrences. If the number mismatches a predefined threshold (e.g., no more than 2 key points for a low-information user), the refiner adjusts the information density accordingly.}

    \item \textcolor{black}{\textit{Formality}: The tool determines and adjusts response formality by illustrative examples in the prompt. The LLM-based judger classifies the formality level, while the refiner modifies responses to match the specified degree of formality.}

    \item \textcolor{black}{\textit{Sentence length}: The tool controls response length by using a rule-based judger to compare it against a predefined threshold (e.g., no more than 20 words for a concise user). If needed, the LLM-powered refiner adjusts the response length while preserving formality and information richness.}

\end{itemize}

The refinement module is highly adaptable, allowing refinement tools to be added or removed as needed. New tools can be created simply by providing in-context examples that demonstrate how to judge and refine responses based on user personas, ensuring precise control over the simulator’s output.

\vspace{-2pt}
\section{Evaluation}
In this section, we evaluate RecUserSim across three key aspects. First, we assess its simulation quality to demonstrate its effectiveness in modeling both individual users and diverse user population. Next, we examine the reliability of its rating mechanism to verify its ability to evaluate different CRS performances. Finally, we present deployment results in an industrial setting, showcasing its effectiveness in real-world CRS evaluation.

\vspace{-2pt}
\subsection{Evaluation on Simulation Quality} \label{cha:exp1}

\textcolor{black}{An ideal user simulator for CRS evaluation should effectively showcase both realistic individual role-playing and the diverse representation of user population. This section presents both subjective and objective evaluations of RecUserSim. Subjectively, we compare it with existing simulators across six dimensions to highlight its strengths in single-user simulation. Objectively, we analyze response diversity to demonstrate its ability to generate diverse output, supporting its effectiveness in simulating large user population. Finally, we highlight RecUserSim's strong controllability over output, underscoring its superiority in individual user simulation.}

\textbf{Baselines.} We evaluate RecUserSim in an unconstrained food recommendation scenario, where recommended restaurants are not limited by real-world availability or user location constraints. \textcolor{black}{To enable comprehensive comparisons, we select baselines representing the two main categories of LLM-based user simulators: single-prompt and agent-based approaches.} Specifically, we compare against:
(1) iEvaLM~\cite{wang-etal-2023-rethinking-evaluation}, a single-prompt simulator that interacts with the CRS using a target item as input; and
(2) CSHI~\cite{zhu2024llmbasedcontrollablescalablehumaninvolved}, an agent-based LLM user simulator incorporating profile, memory, and action modules.
Since existing user simulators were originally designed for movie recommendation tasks, we adapt their prompts and plugin parameters for food recommendation. Additionally, to facilitate interactions with these simulators, we construct a custom \textit{gpt-4o-mini}-based single-prompt CRS.

\textbf{LLM Bases.} The simulators are running on base LLMs including \textit{gpt-3.5-turbo}, \textit{gpt-4o-mini}, and \textit{gpt-4o}\footnote{https://openai.com/index/hello-gpt-4o/}. To demonstrate our user's robustness across different models, we also deploy RecUserSim using the locally hosted \textit{glm-4-9b-chat}~\cite{glm2024chatglm}. 

\subsubsection{Subjective Evaluation of User Simulator Performance}
To thoroughly assess the simulator's ability to emulate a single user, we define six subjective metrics across three dimensions: 

\begin{itemize}[leftmargin=10pt, itemsep=2pt, parsep=0pt, topsep=2pt]
    \item \textit{Single-turn output quality}: Evaluate the simulator's \textbf{naturalness and clarity} in an individual response---whether dialogues flow smoothly and effectively convey information.
    \item \textit{Single-round interaction quality}: Assess the simulator's \textbf{adaptability and relevance} within a single interaction round---the simulator's ability to adjust responses based on CRS actions and provide pertinent feedback.
    \item \textit{Overall dialogue quality}: Measures the simulator's \textbf{role-play ability and realism} throughout the entire dialogue---whether it maintains the user role without offering recommendations and how closely it resembles human interaction patterns.
\end{itemize}

\textbf{Metrics.} \textcolor{black}{We conduct a pairwise comparison between RecUserSim and each baseline model, evaluating their performance across the six subjective metrics. Each simulator interacts with the same CRS to generate 500 dialogues, ensuring a fair comparison.}

\textcolor{black}{For each round, we compare dialogues produced by RecUserSim and a baseline model on all six metrics. A \textit{gpt-4o}-based judge~\cite{2024/LLMasJudge} evaluates these dialogues head-to-head, determining which model performs better for each metric, with possible outcomes of win, draw, or loss for RecUserSim. By aggregating results across 500 comparisons per baseline, we compute RecUserSim's win rate for each metric, quantifying its advantage over baseline simulators.}

\begin{figure}[t]
  \centering
  \includegraphics[width=0.95\columnwidth]{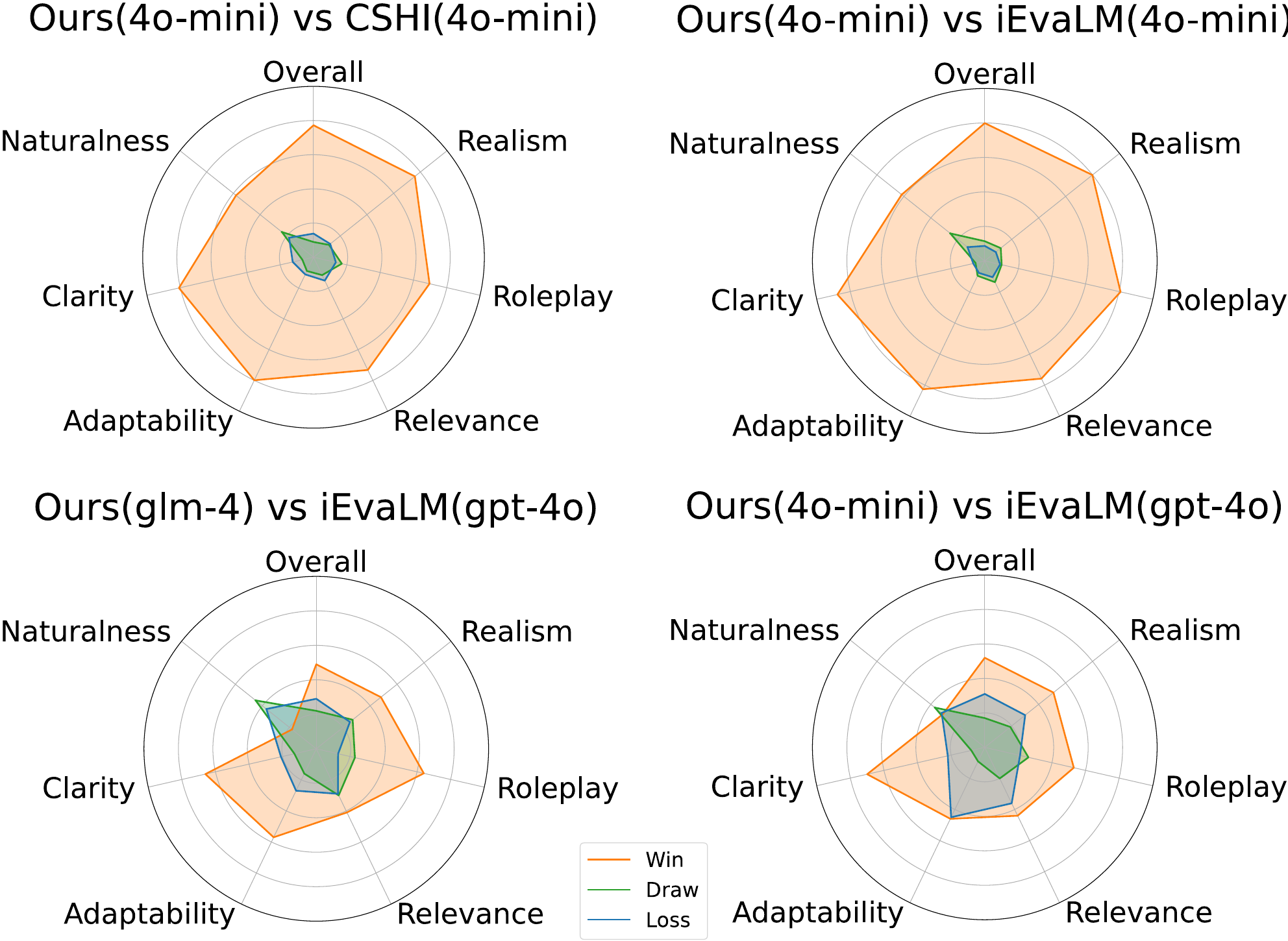}
  \caption{Subjective comparison of RecUserSim and baseline simulators. Higher win rates indicate better performance.}
  % \vspace{-1.5pt}
  \label{fig:subjective compare}  
\end{figure}

\textbf{Results.} As shown in Figure~\ref{fig:subjective compare}, RecUserSim outperforms both iEvaLM and CSHI across all metrics when using the same base LLM, \textit{gpt-4o-mini}, demonstrating its ability to produce higher-quality outputs. \textcolor{black}{Furthermore, even replacing RecUserSim’s base LLM with a smaller open-source model (\textit{glm-4-9b-chat}), it still achieves higher win rates than iEvaLM running on \textit{gpt-4o} in most metrics, except for naturalness. This result highlights RecUserSim's robustness and strong adaptability, confirming that it can maintain high-quality simulation even with a less powerful LLM base.}
\\
\\
\subsubsection{Objective Evaluation of User Simulator Diversity} \label{cha:exp_diver}
\

\textbf{Metrics.} \textcolor{black}{We objectively evaluate the diversity of user simulator outputs by analyzing the distribution of sentence length, information richness, and formality, demonstrating their effectiveness in modeling diverse user population.}

\textcolor{black}{For each model, we generate 500 dialogues and evaluate them based on  sentence length, information richness, and formality. Sentence length is directly measured by word count, while a \textit{gpt-4o}-based evaluator assesses information richness by identifying key points in each sentence and determines formality based on linguistic features. Responses are then categorized using predefined thresholds set by the refinement module.}

\textcolor{black}{By analyzing the distribution of these categories, we evaluate the diversity of simulator outputs. A more uniform distribution indicates greater response diversity, while a skewed distribution suggests a bias towards a specific type, indicating lower diversity.}

\begin{figure}[t]
\centering
\includegraphics[width=0.77\linewidth]{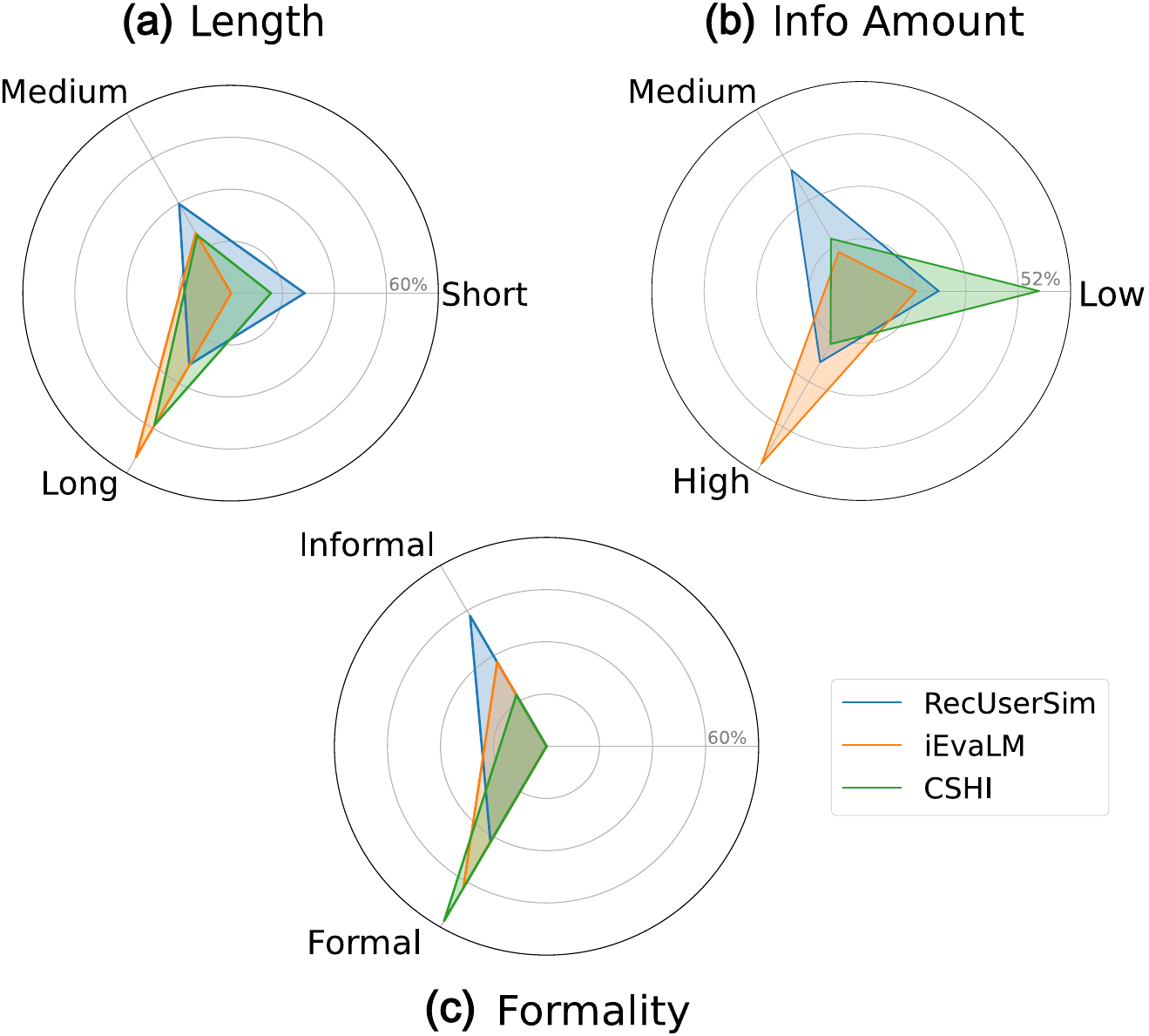}
\caption{Objective comparison of simulators' output diversity. More uniform distributions indicate greater diversity.}
\vspace{-1.5pt}
\label{fig:diversity}
% \vspace{-5pt}
\end{figure}

\textbf{Results.} As shown in Figure~\ref{fig:diversity}, RecUserSim exhibits greater diversity across all metrics compared to iEvaLM and CSHI. It produces a well-balanced mix of sentence lengths, information richness levels, and formality types. In contrast, iEvaLM and CSHI exhibit skewed distributions, with iEvaLM favoring highly informative and formal responses, while CSHI tends to generate formal and less informative replies. These findings confirm RecUserSim’s superior capability in simulating a broad spectrum of user polupations.

\subsubsection{Controllability of RecUserSim over Outputs}

\begin{figure*}[htbp]
\vspace{-1.5pt}
\centering
\includegraphics[width=2.0\columnwidth]{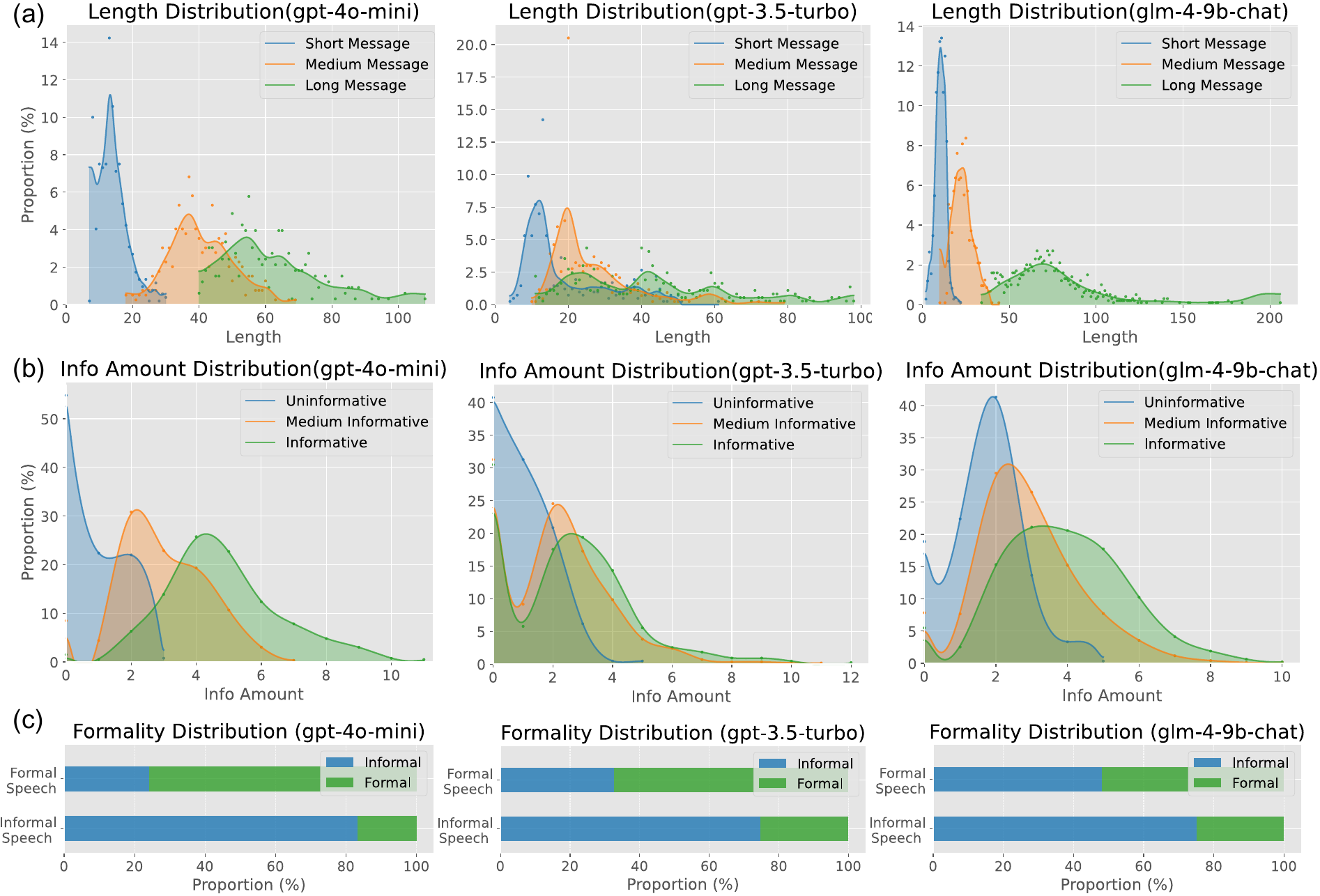}
\vspace{-0.5em}
\caption{Controllability of RecUserSim on outputs based on users' linguistic patterns.}
\vspace{-8pt}
\label{fig:controlability}
\end{figure*}
\

\textbf{Metrics.} \textcolor{black}{We evaluate RecUserSim’s controllability by analyzing how user-specific linguistic patterns influence the distributions of sentence length, information richness, and formality, using the same evaluator as Section~\ref{cha:exp_diver}. A well-controlled simulator should generate outputs that align with users’ linguistic patterns. For instance, users favoring short messages should exhibit distributions skewed toward shorter lengths, while those preferring long messages should produce distributions concentrated at longer outputs.}

\textbf{Results.} Figure~\ref{fig:controlability} illustrates RecUserSim's control over output. Panel (a) shows that users preferring short or long messages generate correspondingly shorter or longer responses, indicating effective length control. Panel (b) illustrates that informative users produce richer information compared to uninformative users. Finally, panel (c) demonstrates control over formality: users inclined towards formal speech yield more formal responses, while those preferring informal speech generate more informal replies. These findings validate RecUserSim’s strong controllability across different base LLMs, reinforcing its advantage in individual user simulation.

\subsection{Evaluation on Rating Reliability} \label{cha:exp2}

An explicit rating mechanism is essential for quantitatively evaluating CRS performance. However, previous work often lacked such a mechanism. To address this, we designed a rating mechanism that simulates the human cognitive process, providing more precise evaluations. To further validate our approach, this section assesses the reliability of our rating mechanism in evaluating CRS with consistency and accuracy across different base LLMs.

\textbf{Baselines.} In the unconstrained food recommendation scenario, we implement two CRS frameworks based on previous work~\cite{li-etal-2024-LLMCRS}: (1) \textbf{BaseCRS}: A single-prompt CRS that generates recommendations based on the dialogue history. (2) \textbf{AgentCRS}: A CRS agent equipped with planning, memory, and action modules. The planning module selects actions; the memory module stores interaction history and extracts user preferences; the action module performs "ask," "recommend," and "answer" actions, each guided by specific prompts. Each framework is implemented with three LLM backbones—\textit{gpt-3.5-turbo}, \textit{gpt-4o-mini}, and \textit{gpt-4o}—yielding six CRS models for evaluation.

\textbf{Metrics.} We employ RecUserSim, built on different base LLMs, to interact with the six CRS models, generating 500 dialogues per pair.
\textcolor{black}{Each CRS is evaluated based on the three dimensions as defined in Section~\ref{cha:action_mecha}: action quality, language quality, and recommendation quality. For each pair, we compute the average score across these 500 dialogues for each dimension, serving as the final model score.}
To measure the consistency of RecUserSim's ratings across different LLM backbones, we compute both the Pearson and Spearman correlation coefficients.

\begin{table*}[t]
    \centering
    \vspace{-1pt}
    \caption{Evaluation results of RecUserSim in assessing CRS models with different base LLMs.}
    \vspace{-0.6em}
    \scalebox{0.95}{
    \begin{tabular*}{\textwidth}{@{\extracolsep{\fill}} llcccccc}
        \toprule
        & & \multicolumn{3}{c}{BaseCRS} & \multicolumn{3}{c}{AgentCRS} \\
        \cmidrule(lr){3-5} \cmidrule(lr){6-8}
        & & gpt-3.5-turbo & gpt-4o-mini & gpt-4o & gpt-3.5-turbo & gpt-4o-mini & gpt-4o \\
        \midrule
        \multirow{3}{*}{RecUserSim (gpt-4o)} & Action & 4.48 & 4.79 & 4.75 & 3.63 & 4.30 & 4.43 \\
        & Language & 4.97 & 4.99 & 4.99 & 4.98 & 4.99 & 5.00 \\
        & Recommendation & 3.96 & 4.01 & 3.97 & 3.76 & 3.98 & 3.98 \\
        \midrule
        \multirow{3}{*}{RecUserSim (gpt-4o-mini)} & Action & 4.35 & 4.73 & 4.77 & 3.66 & 4.31 & 4.42 \\
        & Language & 4.51 & 4.62 & 4.65 & 4.55 & 4.70 & 4.61 \\
        & Recommendation & 3.60 & 3.71 & 3.75 & 3.61 & 3.71 & 3.75 \\
        \midrule
        \multirow{3}{*}{RecUserSim (glm-4-9b)} & Action & 4.06 & 4.39 & 4.32 & 3.59 & 4.11 & 4.18 \\
        & Language & 4.67 & 4.61 & 4.76 & 4.67 & 4.62 & 4.71 \\
        & Recommendation & 3.75 & 3.92 & 3.86 & 3.79 & 3.86 & 3.95 \\
        \bottomrule
    \end{tabular*}}

    \label{tab:evaluation_result}
\vspace{-0.5em}    
\end{table*}

\textbf{Results.} The evaluation results (Table~\ref{tab:evaluation_result}) show that the scores assigned to each CRS align with the performance of their respective base LLMs (\textit{gpt-4o} > \textit{gpt-4o-mini} > \textit{gpt-3.5-turbo}), demonstrating the accuracy of RecUserSim’s evaluations. Despite AgentCRS’s more sophisticated framework, its constrained action space limits behavioral adaptability, leading to lower action ratings than BaseCRS.
Correlation analysis (Tables~\ref{tab:correlation_action} and~\ref{tab:correlation_recommendation}) confirms strong consistency in action ratings (Pearson and Spearman correlations > 0.8) and moderate alignment in recommendation ratings (correlations > 0.5). The slightly lower correlations in recommendation ratings stem from minor score variations among similarly performing models, which are amplified by the sensitivity of correlation measures. Nevertheless, the relatively high correlation still indicates strong agreement between ratings across different LLM backbones, supporting the robustness of RecUserSim’s evaluation mechanism. Overall, these findings validate the accuracy and reliability of RecUserSim in assessing CRS performance across different base LLMs.

\begin{table}[t]
    \centering
    \caption{Correlation values for Action ratings. The $^*$ symbol indicates statistical significance at $p < 0.05$.}
    \vspace{-0.6em}
    \begin{tabular}{lcc}
        \toprule
        Evaluation Models & Pearson & Spearman \\
        \midrule
        gpt-4o vs gpt-4o-mini & $0.99^*$ & $0.88^*$ \\
        gpt-4o vs glm-4-9b & $0.98^*$ & $0.82^*$ \\
        gpt-4o-mini vs glm-4-9b & $0.99^*$ & $0.89^*$ \\
        \bottomrule
    \end{tabular}
    \label{tab:correlation_action}
\vspace{-0.2em}  
\end{table}

\begin{table}[t]
    \centering
    \caption{Correlation value for Recommendation ratings. The $^*$ symbol indicates statistical significance at $p < 0.05$.}
    \vspace{-0.6em}
    \begin{tabular}{lcc}
        \toprule
        Evaluation Models & Pearson & Spearman \\
        \midrule
        gpt-4o vs gpt-4o-mini & $0.62\ $ & $0.51\ $ \\
        gpt-4o vs glm-4-9b & $0.53\ $ & $0.81^*$ \\
        gpt-4o-mini vs glm-4-9b & $0.87^*$ & $0.81^*$ \\
        \bottomrule
    \end{tabular}

    \label{tab:correlation_recommendation}

\vspace{-1em}  
\end{table}

\vspace{-4.5pt}
\subsection{Industrial Deployment}
We have deployed RecUserSim in the development and evaluation of Huawei's Celia Food Assistant, a conversational food recommendation system. In this scenario, the system needs to account for users' geographical locations and restaurant availability to deliver relevant recommendations. We use RecUserSim to evaluate both the demo and online versions of the Celia Food Assistant. Meanwhile, we also conducted human evaluations by randomly selecting 100 conversations from each version. Six evaluators annotated these conversations based on the same criteria of RecUserSim. As shown in Table~\ref{tab:evaluation_result_industry}, RecUserSim's results generally align with those of human evaluators. This consistency demonstrates RecUserSim's effectiveness in evaluating real-world CRS in an industrial setting.

\begin{table}[t]
    \centering
    \caption{Evaluation results of RecUserSim in developing Celia Food Assistant.}
    \vspace{-0.6em}
    \renewcommand{\arraystretch}{1.1}
    \setlength{\tabcolsep}{3pt}
    \begin{tabular}{l >{\raggedright\arraybackslash}p{2.0cm} c c}
        \toprule
        & & Celia (Demo) & Celia (Online) \\
        \midrule
        \multirow{3}{*}{\makecell[c]{RecUserSim \\ (4o-mini/4o)}}
            & Action & 3.37 / 3.30 &3.84 / 3.96 \\
            & Language  & 4.13 / 4.71 & 4.24 / 4.89 \\
            & Recommendation  & 2.36 / 2.04 & 2.99 / 2.59 \\
        \midrule
        \multirow{3}{*}{\ \ \ \ \ Human} 
            & Action & 2.16 & 3.81 \\
            & Language & 4.77 & 4.90 \\
            & Recommendation  & 2.06 & 3.53 \\
        \bottomrule
    \end{tabular}
    \label{tab:evaluation_result_industry}
\vspace{-1.0em}  
\end{table}

\vspace{-4.5pt}
\section{Conclusion}

This paper introduces \textbf{RecUserSim}, a novel LLM agent-based user simulator designed for accurate and comprehensive evaluation of conversational recommender systems. RecUserSim achieves both realistic user role-playing and diverse user population representation through a carefully designed agent framework comprising four key modules.
Specifically, \textcolor{black}{the profile module constructs fine-grained and diverse user personas, laying the foundation for realistic and diverse user simulation. Building upon this, the memory, action, and refinement modules work in tandem to ensure that interactions remain consistent with these personas. This is achieved through mechanisms including unknown preference excitation and three-tier `Rating-Action-Response', which enhance behavioral control and decision-making. Additionally, a tool-augmented refinement method aligns responses with predefined user linguistic patterns, further improving role-playing realism.
To evaluate RecUserSim’s effectiveness, we conducted comprehensive experiments, demonstrating its superiority over existing simulators in generating more realistic, diverse, and high-quality dialogues. Furthermore, its successful deployment in an industrial scenario highlights its practicality and adaptability for real-world applications.}

\vspace{-4.5pt}
\section*{Acknowledgments}
This work is supported in part by National Natural Science Foundation of China (No. 62422215 and No. 62472427), Beijing Outstanding Young Scientist Program NO.BJJWZYJH012019100020098, Intelligent Social Governance Platform, Major Innovation \& Planning Interdisciplinary Platform for the “DoubleFirst Class” Initiative, Renmin University of China, Public Computing Cloud, Renmin University of China, fund for building world-class universities (disciplines) of Renmin University of China, Intelligent Social Governance Platform. This work is also sponsored by Huawei Innovation Research Programs. We gratefully acknowledge the support from Mindspore\footnote{\url{https://www.mindspore.cn}}, CANN (Compute Architecture for Neural Networks) and Ascend AI Processor used for this research.

\bibliographystyle{ACM-Reference-Format}
\bibliography{sample-base}
\end{document}